# Anomalous temperature dependence of current induced torques in CoFeB|MgO heterostructures with Ta based underlayers

Junyeon Kim[1], Jaivardhan Sinha[1], Seiji Mitani[1] and Masamitsu Hayashi[1*]
Saburo Takahashi[2] and Sadamichi Maekawa[3]
Michihiko Yamanouchi[4,5] and Hideo Ohno[4,5,6]

[1]*National Institute for Materials Science, Tsukuba 305-0047, Japan*
[2]*Institute for Materials Research, Tohoku University, Sendai 980-8577, Japan*
[3]*Advanced Science Research Center, Japan Atomic Energy Agency, Tokai 319-1195, Japan*
[4] *Center for Spintronics Integrated Systems, Tohoku University, Sendai 980-8577, Japan*
[5]*Research Institute of Electrical Communication, Tohoku University, Sendai 980-8577, Japan*
[6]*WPI Advanced Institute for Materials Research, Tohoku University, Sendai 980-8577, Japan*

**We have studied the underlayer thickness and temperature dependences of the current induced effective field in CoFeB|MgO heterostructures with Ta based underlayers. The underlayer thickness at which the effective field saturates is found to be different between the two orthogonal components of the effective field, i.e. the damping-like term tends to saturate at smaller underlayer thickness than the field-like term. For large underlayer thickness films in which the effective field saturates, we find that the temperature significantly influences the size of the effective field. A striking difference is found in the temperature dependence of the two components: the damping-like term decreases whereas the field-like term increases with increasing temperature. Using a simple spin diffusion-spin transfer model, we find that all of these results can be accounted for provided the real and imaginary parts of an effective spin mixing conductance are negative. These results imply that either spin transport in this system is different from conventional metallic interfaces or effects other spin diffusion into the magnetic layer need to be taken account in order to model the system accurately.**

*Email: hayashi.masamitsu@nims.go.jp

Materials and/or interfaces with large spin orbit coupling are attracting considerable interest since they can generate significant amount of spin current or accumulate spins to manipulate magnetic moments[1, 2]. The spin Hall effect[3-5] and current induced spin polarization (i.e. the Rashba-Edelstein effect) are considered to be the major sources that enable spin current generation and spin accumulation, respectively, in ultrathin magnetic heterostructures. The non-equilibrium spins can act on the magnetic moments via spin transfer torque[6, 7] or exchange torque[8-10] to trigger magnetization switching. These torques are termed the "spin orbit torques" which are to be distinguished from the conventional spin transfer torque since the system requires strong spin orbit coupling to generate the spins.

The action of spins on the magnetic moments can be evaluated by measuring the "effective magnetic field" which reflects the size and direction of the spin orbit torque. Recently, it has been shown[11-14] that the spin orbit torque also possesses a damping-like torque[6, 7] and a field-like torque[15], analogous to the spin transfer torque in magnetic tunnel junctions[16-20]. In many systems, however, the field-like torque is much larger than the damping-like torque[11, 14], and its direction is pointing opposite to the incoming spin direction of the electrons (assuming the spin source is the spin Hall effect) [11, 14, 21]. Moreover, the size and direction of both components of the torque vary depending on the materials and the film structures[11-14]. Up to date, many of these results cannot be accounted for with existing theories that consider spin transfer and/or exchange torques[22, 23]. It is thus essential to understand the origin of spin orbit torques in order to apply them for possible applications in Spintronic devices.

Here we study the spin orbit torque in CoFeB|MgO heterostructures with Ta based underlayers as a function of the underlayer thickness and temperature to provide insight into spin transport and action of non-equilibrium spins on the magnetization. The underlayer thickness at which the effective field saturates differs between the damping-like and field-like terms. Surprisingly, for a thicker underlayer films, we find that the two orthogonal

components of the effective field show opposite temperature dependence; the field-like term increases and the damping-like term decreases with increasing temperature. These results can be accounted for with a simple spin diffusion-spin transfer model provided we use a rather unconventional value of spin mixing conductance.

Films are deposited by magnetron sputtering on Si substrates coated with ~100 nm thick thermally oxidized Si. The film structures studied are A: Sub.|$d$ Ta|1 CoFeB|2 MgO|1 Ta and B: Sub.|$d$ TaN|1 CoFeB|2 MgO|1 Ta (units in nanometer). In our previous study, we find that proper amount of nitrogen introduced into the Ta underlayer increases the film's perpendicular magnetic anisotropy (PMA) and reduces the magnetic dead layer thickness[24]. Whereas the PMA decreases with the Ta underlayer thickness and drops to near zero beyond $d$~1.5 nm for film structure A, it is almost independent of the TaN underlayer thickness, at least till $d$~10 nm, for film structure B. Film structure B thus allows study of the current induced effective field for thicker underlayer films.

Figure 1(a) shows a schematic illustration of the Hall bars patterned from the films along with the definition of the xyz axes. Adiabatic harmonic Hall voltage measurements[25] are used to evaluate the current induced effective field. Briefly, we apply a sinusoidal excitation voltage (frequency: ~507.32 Hz) of amplitude $V_{IN}$ to the Hall bar and measure the first ($V_\omega$) and second ($V_{2\omega}$) harmonic Hall voltages as a function of in-plane magnetic field directed along ($H_L$) or transverse ($H_T$) to the current flow. The two components of the effective field, parallel ($\Delta H_L$) and orthogonal ($\Delta H_T$) to the current flow, are estimated using the following relationships[11]: $\Delta H_{T(L)} \sim -2\left(\frac{\partial V_{2\omega}}{\partial H_{T(L)}}\right) \Big/ \left(\frac{\partial^2 V_\omega}{\partial H_{T(L)}^2}\right)$. Contributions from the planar Hall effect (PHE) on the harmonic voltage signals are taken into account when obtaining $\Delta H_{T(L)}$[12,26]. For the transverse effective field ($H_T$), we calculate the Oersted field generated by the current flowing through the Ta(N) underlayer and subtract its contribution.

Figure 1(b-e) show the temperature dependence of the longitudinal resistance ($R_{XX}$), the anomalous Hall effect (AHE) resistance ($\Delta R_{XY}$), the saturation magnetization ($M_S$) and the effective magnetic anisotropy energy ($K_{EFF}$) for the Ta and TaN underlayer films. All quantities are normalized at their room temperature value (see supplementary information for the corresponding absolute values). $\Delta R_{XY}$ corresponds to one half of the change in the anomalous Hall resistance ($R_{XY}$) when the magnetization of the CoFeB layer is reversed. Positive $K_{EFF}$ indicates that the magnetic easy axis points along the film normal.

For all film structures investigated, $R_{XX}$ shows little dependence on the temperature in the range studied here (~100-360 K): the variation is less than 10%. This is likely to do with the amorphous-like texture of the Ta and CoFeB layers[27]. Note that $R_{XX}$ reflects changes in the resistivity of both the underlayer and the CoFeB layer. From separate measurements, we find that the temperature variation of the underlayer resistivity is small and is similar to that shown in Fig. 1(b). $\Delta R_{XY}$ displays larger temperature dependence than $R_{XX}$, in particular, for the films with thicker Ta underlayer. Interfacial contributions to the AHE may be responsible for the large temperature variation of $\Delta R_{XY}$[28, 29]. The change in $K_{EFF}$ with temperature is larger than that of $M_S$; we find a relationship $K_U \propto M_S^n$ [30] with $n$ ranging from ~2.2-2.7 where $K_U = K_{EFF} + 2\pi M_S^2$; see the inset of Fig. 1(e). Within the small temperature range studied here, it is difficult to provide accurate values of $n$ since the change in $M_S$ is rather small (and we have assumed the forth order anisotropy constants to be negligible).

Both transverse ($\Delta H_T$) and longitudinal ($\Delta H_L$) components of the current induced effective field scale linearly with the input voltage ($V_{IN}$) at low excitation. We fit $\Delta H_{T(L)}$ vs. $V_{IN}$ with a linear function to obtain the effective field per unit excitation. $V_{IN}$ can be converted to the current density flowing through the underlayer ($J_N$) by taking into account the resistivity difference between the underlayer and the CoFeB layer. Solid and open symbols in Fig. 2 show $\Delta H_{T(L)}/J_N$ of the Ta underlayer films for equilibrium magnetization direction ($M_Z$) pointing along +z and −z, respectively. $\Delta H_L$ changes its sign upon reversing the magnetization direction, whereas $\Delta H_T$ is independent of $Mz$, which are characteristics of the damping-like and field-like terms, respectively. Note that the definition of the coordinate system (xyz axes) is different from our previous report: negative $\Delta H_T$ here corresponds to positive $\Delta H_T$ in Ref. [11] (sign of $\Delta H_L$ is the same). Both components display significant temperature dependence. The field-like term ($\Delta H_T$) increases with increasing temperature for films with thick Ta layer: the curves can be fitted with a growing exponential function. Interestingly, $\Delta H_T$ reduces to near zero at temperatures of ~100 K. In contrast, the damping-like term ($\Delta H_L$) decreases as the temperature increases: it tends to saturate at lower temperature.

Previously, we have reported that both $\Delta H_T$ and $\Delta H_L$ change their sign at a certain Ta underlayer thickness ($d$)[11]. We have assumed that there are two competing sources that contribute to the generation of the effective field, one likely the spin Hall effect and the spin transfer torque (referred to as the "spin Hall torque" from hereafter) which depends on $d$, and the other an interfacial effect (e.g. the Rashba-Edelstein effect) that shows little dependence on $d$[11]. The two effects generate $\Delta H_{T(L)}$ that point against each other at room temperature. Here we find that the sign change also takes place with temperature for a given Ta thickness. Figure 2 shows that the spin Hall torque displays large temperature dependence whereas the interfacial torque is rather insensitive to temperature (see e.g. $\Delta H_T$ for $d$~0.1 and 0.3 nm).

It is generally assumed that the spin Hall torque generates both the damping-like and the field-like components. However, the difference in the temperature dependence of $\Delta H_T$ and $\Delta H_L$ suggests that instead the origin of the two components may be different. We thus use a simplified model to examine whether the different temperature dependence can be explained by a single effect, i.e. the spin Hall torque. To model the system, we consider a bilayer consisting of a heavy non-magnetic metal (NM) underlayer and a ferromagnetic (FM) layer and assume that the current induced effective field ($\Delta H$) scales with an interface spin current (at the NM|FM interface) governed by the spin mixing conductance $G_{MIX}$[31, 32]. The two orthogonal components of $\Delta H$ can be expressed as (see supplementary information)[33]:

$$\Delta H_x = -\theta_{SH} \frac{\hbar}{2|e|} \frac{J_N}{M_S t_F} \left(1 - \frac{1}{\cosh(d/\lambda_N)}\right) \left[\frac{(1+g_r)g_r + g_i^2}{(1+g_r)^2 + g_i^2}\right] \text{sgn}(\hat{m} \times \hat{e}_y)_x \quad (1)$$

$$\Delta H_y = -\theta_{SH} \frac{\hbar}{2|e|} \frac{J_N}{M_S t_F} \left(1 - \frac{1}{\cosh(d/\lambda_N)}\right) \left[\frac{g_i}{(1+g_r)^2 + g_i^2}\right] \quad (2)$$

where $e$ and $\hbar$ are the electric charge and the Planck constant, $d$, $\rho_N$, $\lambda_N$ and $\theta_{SH}$ are the thickness, resistivity, spin diffusion length and the spin Hall angle of the NM layer, $t_F$ and $M_S$ are the thickness and saturation magnetization of the FM layer. We define $g_r = \text{Re}[G_{MIX}]\rho_N \lambda_N \coth\left(\frac{d}{\lambda_N}\right)$ and $g_i = \text{Im}[G_{MIX}]\rho_N \lambda_N \coth\left(\frac{d}{\lambda_N}\right)$, where $\text{Re}[G_{MIX}]$ and $\text{Im}[G_{MIX}]$ are the real and imaginary parts of $G_{MIX}$. $\hat{m}$ and $\hat{e}_y$ are unit vectors representing the magnetization direction and +y. $\Delta H_x$ and $\Delta H_y$ correspond to the damping-like ($\Delta H_L$) and the field-like ($\Delta H_T$) terms, respectively. Contribution from the spins which enter the FM layer is neglected in obtaining $\Delta H$ (the effect of spin transport within the FM layer can be included effectively in $G_{MIX}$).

To test the validity of Eqs. (1) and (2), we first fit the underlayer thickness ($d$) dependence of $\Delta H_i$ for the TaN underlayer films, as shown in Fig. 3. (From hereafter we focus on the results from the TaN underlayer films since the range of $d$ allowed to study the effective field is larger than that of the Ta underlayer films.) There are total 6 material parameters in Eq. (1) and (2) that need to be determined. Since $M_S$ and $\rho_N$ vary little with temperature, for simplicity, we assume that they are temperature independent and we use their room temperature value: $M_S$=1200 emu/cm$^3$ and $\rho_N$=395 μΩ·cm. Similarly, we assume the spin diffusion length ($\lambda_N$) and the spin Hall angle ($\theta_{SH}$) to be temperature independent given the little variation of $\rho_N$ with temperature. The remaining parameters, Re[$G_{MIX}$] and Im[$G_{MIX}$], are allowed to vary with temperature since we consider $G_{MIX}$ as an *effective* mixing conductance that takes into account not only conventional spin transport at the NM|FM interface but also other unconventional interface effects, perhaps related to the presence of spin orbit coupling.

For a given $\lambda_N$ and $\theta_{SH}$, we fit the underlayer thickness ($d$) dependence of the effective field, measured at 100 K and at room temperature (RT), with Re[$G_{MIX}$] and Im[$G_{MIX}$] as the fitting parameters. The results are shown by the solid lines in Fig. 3, which agree well with the experimental results (here $\lambda_N$=2.5 nm and $\theta_{SH}$=−0.08 are used: see supplementary information for the fitting details). In particular, we find, experimentally, that saturation of $\Delta H_L$ occurs at smaller $d$ than that of $\Delta H_T$ for many films. For example, $\Delta H_L$ saturates at ~5 nm at ~295 K but $\Delta H_T$ increases with $d$ even for the maximum $d$ used. Such feature can be described using Eqs. (1) and (2) for a given $\lambda_N$. Values of Re[$G_{MIX}$] and Im[$G_{MIX}$] used at 100 K and at RT are shown by the solid symbols in Fig. 4(a): we find that both real and imaginary parts of $G_{MIX}$ need to be negative in order to reproduce these results.

The temperature dependence of $\Delta H_T$ and $\Delta H_L$, plotted in Fig. 4(b) and (c) (symbols) for the thickest TaN underlayer film ($d$~8.9 nm), can be reproduced using the parameters found above. Note that temperature dependence for the TaN underlayer films is similar to that of the thicker ($d$>~0.7 nm) Ta underlayer films shown in Fig. 2. To estimate $G_{MIX}$ at each temperature, here we assume for simplicity that $G_{MIX}$ varies linearly with temperature from 100 K to RT (there is no justification for this). At each temperature, we substitute the temperature independent parameters ($M_S$, $\rho_N$, $\lambda_N$ and $\theta_{SH}$) and the interpolated values of Re[$G_{MIX}$] and Im[$G_{MIX}$], as shown by the solid lines in Fig. 4(a), in Eq. (1) and (2). The solid lines in Fig. 4(b) and 4(c) show the calculated results. The trend agrees well with the experimental data. Thus a temperature dependent $G_{MIX}$ with negative real and imaginary parts can describe both the underlayer thickness dependence and the anomalous temperature dependence of the effective field. It should be noted that the size of the real and imaginary parts of $G_{MIX}$ vary depending on the values of $\theta_{SH}$ and $\lambda_N$ used: both Re[$G_{MIX}$] and Im[$G_{MIX}$] lie in a range between ~$-2 \times 10^{10}$ $\Omega^{-1}$cm$^{-2}$ to ~0 $\Omega^{-1}$cm$^{-2}$ in the temperature range studied (see supplementary information).

Previously, the imaginary part of $G_{MIX}$ has been reported to be small and negligible in metallic heterostructures, e.g. in giant magneto-resistive spin valves[16, 34, 35]. The presence of an imaginary part comparable in size with its real counterpart suggests that the spin current reflected at the NM|FM interface experience a large angle rotation of its spin direction. It is generally understood that, however, the net rotation averages out to zero since the precession phase and frequency vary among electrons with different wave vectors[34, 35]. The large imaginary part of $G_{MIX}$ can thus be caused by a different mechanism, perhaps involving a strong spin orbit coupling at interface(s) that has not been considered thus far.

The *negative* Re[$G_{MIX}$] found here is an unusual characteristic compared to conventional spin torque devices. In a free electron model[34], Re[$G_{MIX}$] can be negative if the band matching between the NM and the FM layers is poor and if one allows interband transition(s) upon reflection at the interface. The absolute values of $G_{MIX}$ (~$10^{10}$ $\Omega^{-1}$cm$^{-2}$, see Fig. 4(a)) are indeed smaller than what have been reported recently in other systems[36-38], which is consistent with an interface with large band mismatch. Alternatively, negative Re[$G_{MIX}$] is possible if spin current flows from the FM layer to the NM layer (assuming that the spin Hall torque takes place at the NM|FM interface). One can consider the anomalous Hall effect and/or the spin Hall effect, if any, in the FM layer to provide such source, although its size and sign need to be examined carefully.

It remains to be seen whether the model we use here accurately describes the spin orbit torques in magnetic heterostructures. The unusual values of the spin mixing conductance may change if we explicitly take into account spin transport in the FM layer. In particular, the effect of the other FM layer interface (here, CoFeB|MgO, which provides the perpendicular magnetic anisotropy of the system) on the effective field is not clear at the moment. Other possibilities, such as the Rashba effect with the exchange torque at interface(s)[1, 12] and interfacial magnon excitations[19, 39-41] which are not treated here, may also partly account for the effective field generation and its temperature dependence.

In summary, we have studied the underlayer thickness and temperature dependences of the current induced effective field in perpendicularly magnetized CoFeB|MgO heterostructures with Ta based underlayers. The underlayer thickness at which the effective field saturates differs between the damping-like and the field-like terms. When the underlayer thickness is large so that the two terms saturate, we find that the damping-like component decreases whereas the field-like component increases with increasing temperature. These results can be explained with a model that considers spin current generation via the spin Hall effect and spin transfer torque taking place at the underlayer|CoFeB layer interface, however, with a rather unconventional value of a temperature dependent effective spin mixing conductance. Our results imply that either spin transport at the underlayer|CoFeB interface is different from conventional metallic interfaces (e.g Cu|Co), or effects other than spin diffusion into the magnetic layers, in particular, contributions from the CoFeB|MgO interface, magnon

excitations and Rasha-like effects need to be considered in the model to accurately describe spin orbit torques in ultrathin magnetic heterostructures.

**Acknowledgements**

This work was partly supported by the Grant-in-Aid (25706017) from MEXT and the FIRST program from JSPS.

**Figure captions**

**Figure 1.** (a) Schematic illustration of the experimental setup. Definition of the coordinate system is included. (b-e) Temperature dependence of the longitudinal resistance $R_{XX}$ (b), the anomalous Hall effect resistance $\Delta R_{XY}$ (c), the saturation magnetization $M_S$ (d) and the effective magnetic anisotropy energy $K_{EFF}$ (e), all normalized by their room temperature value for the Ta and TaN underlayer films. Thickness of the underlayer is indicated. Inset of (e) shows $\log_{10}[K_U(T)/K_U(100K)]$ vs. $\log_{10}[M_S(T)/M_S(100K)]$ to obtain the exponent $n$ of $K_U$ vs. $M_S^n$.

**Figure 2.** (a-b) Temperature dependence of $\Delta H_T/J_N$ (a) and $\Delta H_L/J_N$ (b) for the Ta underlayer film. Thickness of the Ta underlayer is indicated. Bottom panels show a magnified view of the upper panel for the thinner Ta underlayer devices. Solid and open symbols correspond to magnetization pointing along +z and -z, respectively.

**Figure 3.** (a-d) Underlayer thickness dependence of $\Delta H_T/J_N$ (a,b) and $\Delta H_L/J_N$ (c,d) for the TaN underlayer film. The measurement temperature is 100 K for (a,c) and room temperature (~295 K) for (b,d). Solid lines are calculated using Eqs. (1) and (2) with the following parameters: $M_S$=1200 emu/cm³, $\rho_N$=395 μΩ·cm, $\lambda_N$=2.5 nm, $\theta_{SH}$=−0.08 and the real and imaginary parts of the spin mixing conductance ($G_{MIX}$), which are shown by the solid symbols in Fig. 4(a).

**Figure 4.** (a) Temperature dependence of the real and imaginary parts of the spin mixing conductance ($G_{MIX}$) obtained by the fitting shown in Fig. 3. Linear fit to the data (symbols) is shown by the solid line. (b-c) Temperature dependence of $\Delta H_T/J_N$ (b) and $\Delta H_L/J_N$ (c) for the TaN underlayer film with $d\sim 8.9$ nm. Solid lines are calculated using Eqs. (1) and (2) with the following parameters: $M_S$=1200 emu/cm$^3$, $\rho_N$=395 μΩ·cm, $\lambda_N$=2.5 nm, $\theta_{SH}$=−0.08 and the interpolated real and imaginary parts of the spin mixing conductance ($G_{MIX}$) shown by the solid lines in Fig. 4(a).

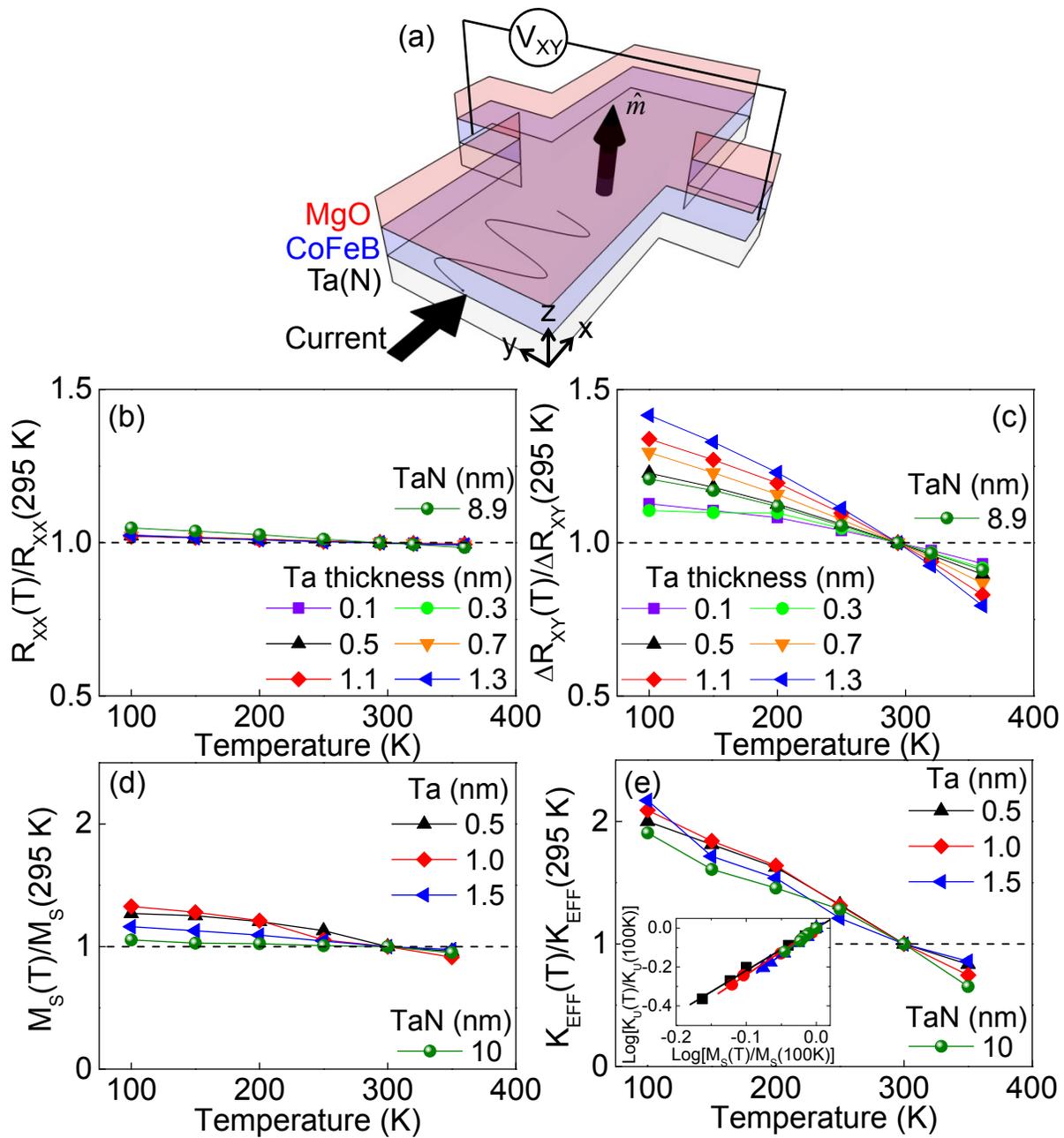

Fig. 1

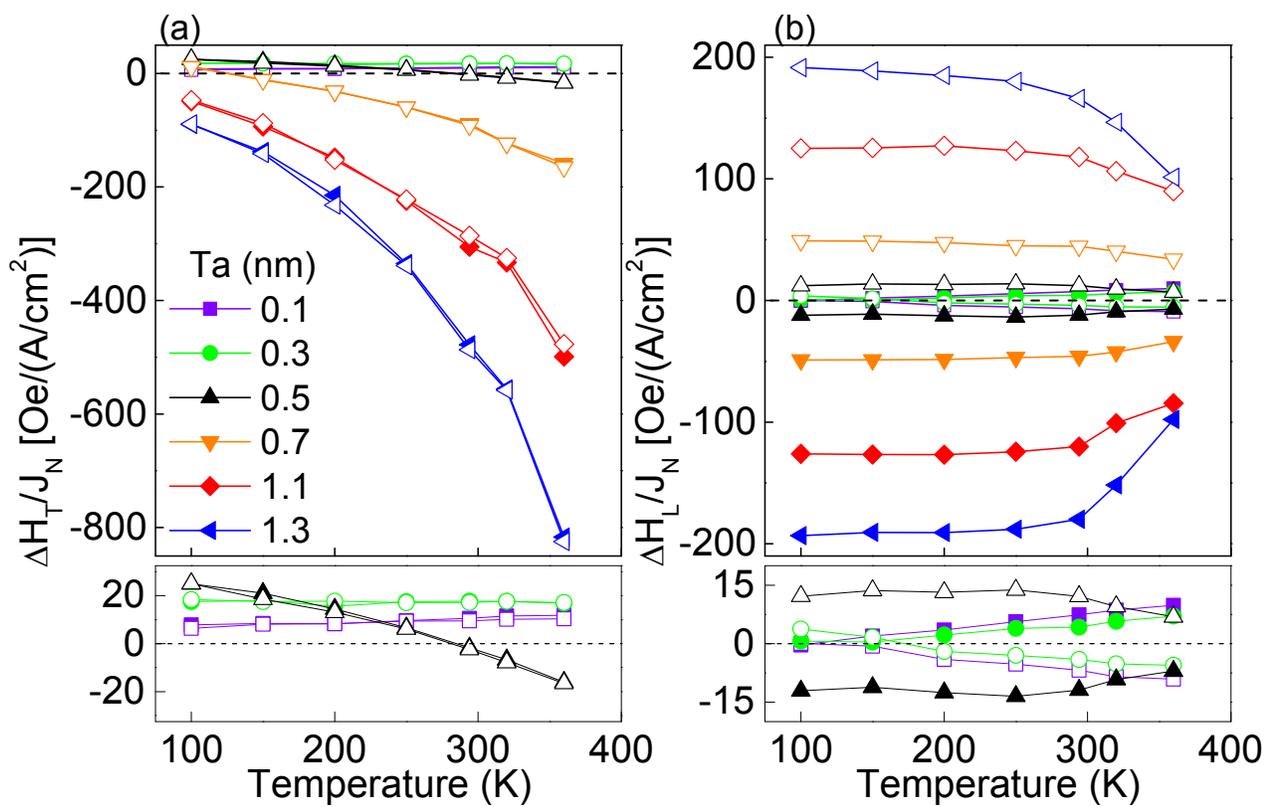

Fig. 2

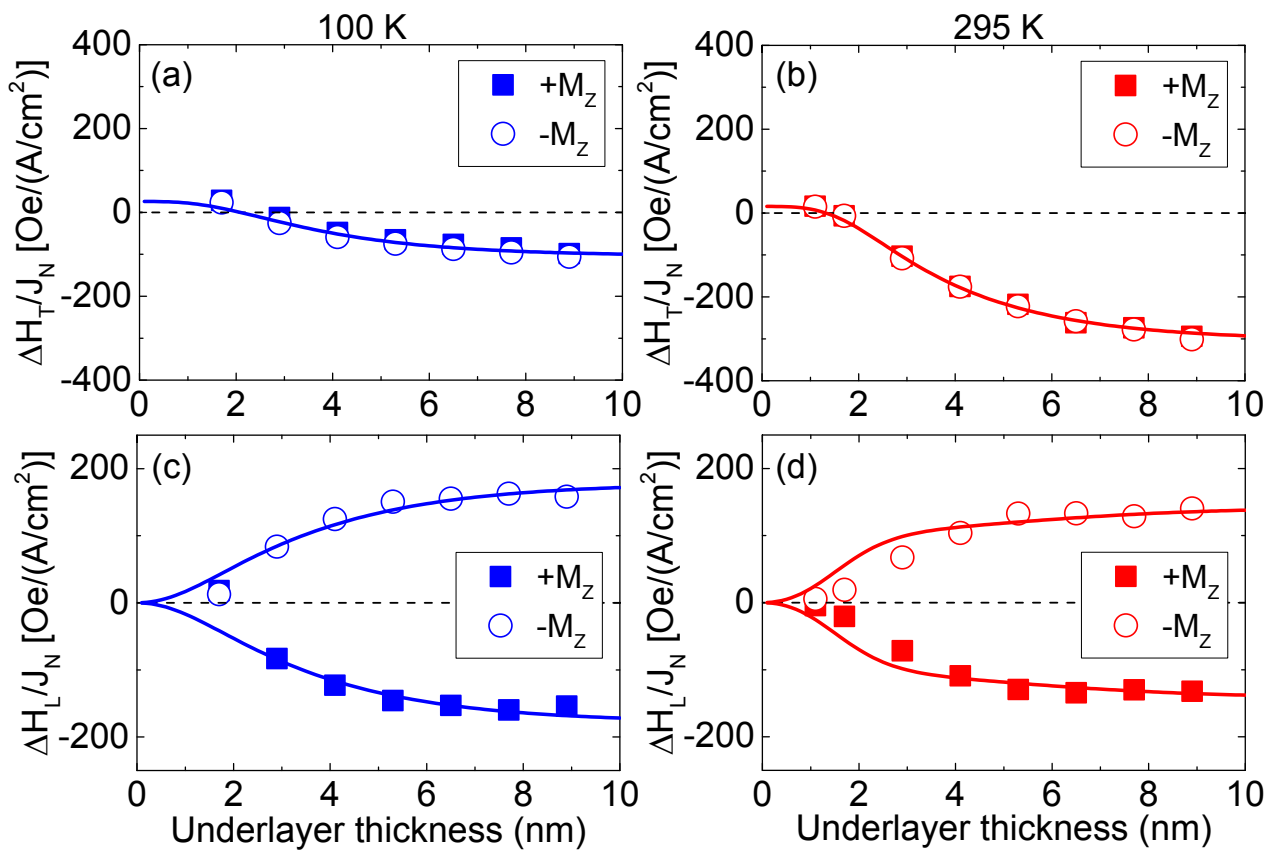

Fig. 3

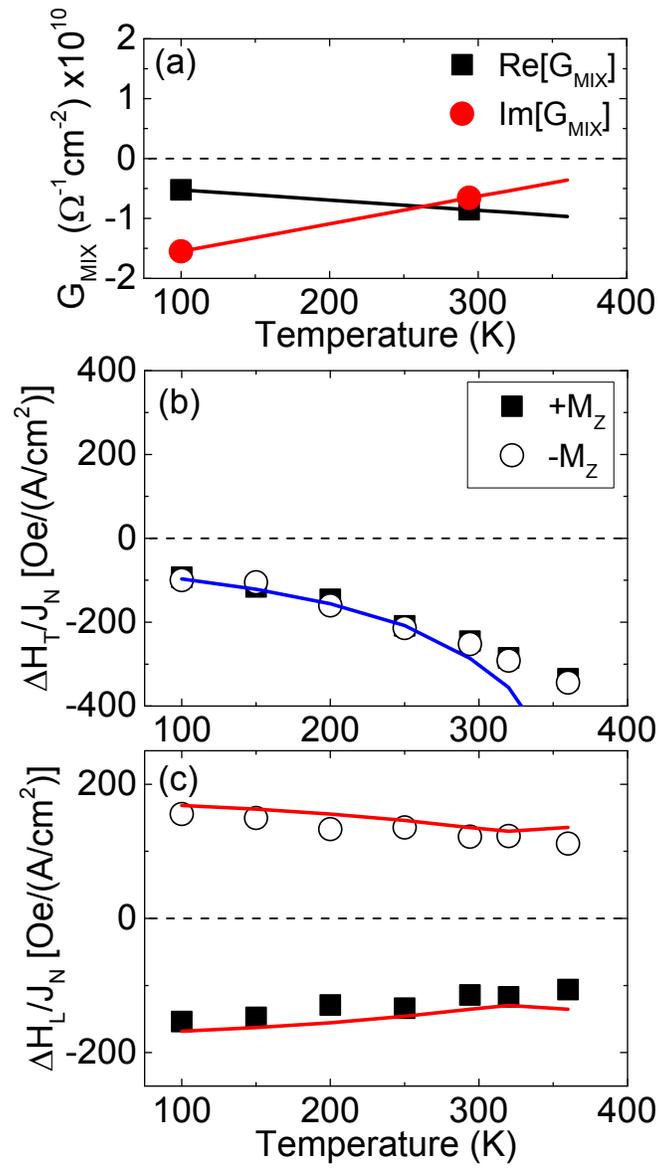

Fig. 4

**Supplementary information for:**

**Anomalous temperature dependence of current induced torques in magnetic heterostructures**

Junyeon Kim[1], Jaivardhan Sinha[1], Seiji Mitani[1] and Masamitsu Hayashi[1*]
Saburo Takahashi[2] and Sadamichi Maekawa[3]
Michihiko Yamanouchi[4,5] and Hideo Ohno[4,5,6]

[1]*National Institute for Materials Science, Tsukuba 305-0047, Japan*
[2]*Institute for Materials Research, Tohoku University, Sendai 980-8577, Japan*
[3]*Advanced Science Research Center, Japan Atomic Energy Agency, Tokai 319-1195, Japan*
[4] *Center for Spintronics Integrated Systems, Tohoku University, Sendai 980-8577, Japan*
[5]*Research Institute of Electrical Communication, Tohoku University, Sendai 980-8577, Japan*
[6]*WPI Advanced Institute for Materials Research, Tohoku University, Sendai 980-8577, Japan*

## S1. Sample preparation and the magnetic/transport properties of Ta(N)|CoFeB|MgO

The Hall bar is patterned from A: Sub.|$d$ Ta|1 CoFeB|2 MgO|1 Ta and B: Sub.|$d$ TaN|1 CoFeB|2 MgO|1 Ta (units in nm) deposited on thermally oxidized (100 nm thick SiO$_2$) Si substrates. TaN is formed by reactively sputtering Ta in Ar gas atmosphere mixed with a small amount of N$_2$ gas (see Ref. [1] for the details). Ar and N$_2$ gas concentrations are controlled independently by gas mass flow meters. We define $Q$ as the atomic ratio of the N$_2$ gas over the total (Ar + N$_2$) gas, i.e. $Q \equiv \frac{S_{N_2}}{S_{Ar} + S_{N_2}}$, where $S_X$ denotes the mass flow (in unit of sccm) of gas X. $Q$ is fixed to ~1.2% here. The resulting atomic composition is Ta$_{44\pm5}$N$_{56\pm5}$ which is determined by Rutherford backscattering spectroscopy (RBS). Films are deposited using a linear shutter to vary the underlayer thickness across the wafer. All films are post-annealed at 300 °C for one hour in vacuum.

Conventional optical lithography and Ar ion etching are used to pattern the Hall bar. Electrodes made of 10 Ta|100 Au (units in nm) are subsequently patterned using a lift-off process. The width ($w$) of the Hall bar is varied from ~5 μm to 20 μm, the length ($L$) is fixed to 60 μm. Sample preparation is described in detail in Ref. [2]. Most of the results shown here are from ~10 and ~20 μm wide Hall bars. The current induced effective field shows little dependence on the Hall bar width.

The longitudinal resistance $R_{XX}$ normalized by a geometrical factor ($L/w$), the Hall resistance $\Delta R_{XY}$, the saturation magnetization $M_S$ and the effective anisotropy energy $K_{EFF}$ are plotted as a function of temperature in Fig. S1. The longitudinal and Hall resistances are evaluated using the patterned Hall bars, whereas $M_S$ are $K_{EFF}$ are measured using continuous films. $M_S$ are $K_{EFF}$ are obtained using Vibrating sample magnetometer (VSM) and Superconducting Quantum Interference Device (SQUID).

## S2. Current induced effective field measurements

To estimate the current induced effective field, the first ($V_\omega$) and second ($V_{2\omega}$) harmonic Hall voltages are measured as a function of in-plane external magnetic field. The effective field components directed along ($\Delta H_L$) and transverse to ($\Delta H_T$) the current flow are obtained by measuring $V_\omega$ and $V_{2\omega}$ with external field pointing along the x- ($H_L$) and y-axis ($H_T$), respectively. Exemplary plots of the first and second harmonic voltages vs. the external in-plane field are shown in Fig. S2(a) and S2(b).

To account for the planar Hall effect contribution, the effective field is evaluated using the following relation[3]:

$$\Delta H_L = -2 \frac{(B_L \pm 2\xi B_T)}{1 - 4\xi^2}$$
$$\Delta H_T = -2 \frac{(B_T \pm 2\xi B_L)}{1 - 4\xi^2}$$
(S2.1)

where $B_L \equiv \frac{\partial V_{2\omega}}{\partial H_L} \Big/ \frac{\partial^2 V_\omega}{\partial H_L^2}$, $B_T \equiv \frac{\partial V_{2\omega}}{\partial H_T} \Big/ \frac{\partial^2 V_\omega}{\partial H_T^2}$ and $\xi \equiv \frac{\Delta R_P}{\Delta R_A}$. $\Delta R_P$ and $\Delta R_A$ are $\Delta R_{XY}$ due to the planar Hall effect and the anomalous Hall effect, respectively. When the planar Hall contribution to $\Delta R_{XY}$ is small, i.e. $\xi \ll 1$, then $\Delta H_{T(L)} = B_{T(L)}$, which is equivalent to the expression shown in the main text. The excitation voltage amplitude ($V_{IN}$) dependence of $\Delta H_T$ and $\Delta H_L$ obtained using the data shown in Figs. S2(a,b) and Eq. (S2.1) are shown in Fig. S2(c) and S2(d).

The temperature and underlayer thickness dependences of the anomalous Hall effect ($\Delta R_A$) and the planar Hall effect ($\Delta R_P$) are measured using the setup illustrated in Fig. S3(a) and S3(b), respectively. For evaluating $\Delta R_A$, the magnitude of the external field is kept constant (~20-30 kOe) and the sample is rotated in the yz plane (Fig. S3(a)). $\Delta R_P$ is measured in a similar way, however, with the sample rotation now in the xy plane (Fig. S3(b)). Note that contribution from the ordinary Hall resistance is small at this field range. $\Delta R_{XY}$ shown in Figs. 1(c) and S1(b) corresponds to $\Delta R_A$ evaluated at low magnetic field (~2 kOe).

The angular dependence of the Hall resistance ($R_{XY}$) are shown in Figs. S3(c) and S3(d) for the two sweep planes for a Hall bar for film structure A: Sub.|$d$ Ta|1 CoFeB|2 MgO|1 Ta with $d$~1.0 nm. The temperature dependence of $\Delta R_P$, $\Delta R_A$ and the ratio $\xi \equiv \dfrac{\Delta R_P}{\Delta R_A}$ evaluated for this device are shown in Figs. S3(e) and S3(f), respectively. The temperature dependence of the ratio $\xi$ for four devices with different $d$ (film structure A) are presented in Fig. S4. The planar Hall contribution to $\Delta R_{XY}$, i.e. $\xi$, is ~5% for most of devices studied.

Since the variation of $\xi$ is small among devices with various $d$, we take the average value of $\xi$ from the four devices measured (as shown in Fig. S4) at each temperature and use Eq. S(2.1) to estimate the effective field; $\xi$ is thus assumed constant with $d$ but varies with temperature according to the results shown in Fig. S4. The effective field components ($\Delta H_{T(L)}$) shown in Figs. 2-4 are evaluated using Eq. (S2.1) with the temperature dependent $\xi$.

The total magnitude and direction of the current induced effective field are shown in Figs. S5(a) and S5(b), respectively, as a function of temperature for both film structures A and B. The magnitude is defined as $|\Delta H / J_N| \equiv \sqrt{(\Delta H_T / J_N)^2 + (\Delta H_L / J_N)^2}$ and the direction is set with respect to the current flow direction, i.e. $\varphi = \arctan\left[\dfrac{\Delta H_T / J_N}{\Delta H_L / J_N}\right]$. In evaluating $\varphi$, we use the effective field ($\Delta H_T/J_N$ and $\Delta H_L/J_N$) for magnetization point pointing along +z. The magnitude increases with temperature due to the presence of large field-like term $\Delta H_T/J_N$ for the thick underlayer films. The angle varies continuously with temperature for all films, and in many cases, it approaches the direction defined by $\Delta H_T/J_N$ at higher temperatures.

**S3. The bilayer model for spin transport**

A. Derivation of Eq. (1) and (2)

We consider a bilayer consisting of a heavy non-magnetic metal (NM) underlayer and a ferromagnetic (FM) layer. A spin current $Q_{ij}$ (left index: spin direction, right index: flow in real space[4]), generated in the NM layer via the spin Hall effect and diffuses into the FM layer[5], is expressed as:

$$Q_{iz}(z) = -\frac{1}{2|e|\rho_N}\frac{\partial}{\partial z}\delta\mu_i + \theta_{SH}\left(\hat{e}_i \times \vec{J}_N\right)_z \tag{S3.1}$$

where $\rho_N$ and $\theta_{SH}$ are the resistivity and the spin Hall angle of the NM layer, $e$ is the electric charge and $\delta\vec{\mu}$ represents the spin accumulation that takes place at the NM|FM interface (within the NM layer). $\hat{e}_i$ ($i$=x,y,z) is an unit vector that represents the spin direction of electrons that drift due to the spin Hall effect. For the geometry described in Fig. 1(a), $\vec{J}_N = J_N\hat{x}$ and the spin direction of the relevant electrons here is $\hat{e}_y$. Boundary conditions for the spin current in the NM layer are assumed as:

$$\begin{aligned}Q_{iz}(z=0) &= 0 \\ Q_{iz}(z=d) &= Q_{iz}^{N|F}\end{aligned} \tag{S3.2}$$

Here z=0 corresponds to the interface between the NM layer and the substrate (100 nm thick SiO$_2$) and $Q_{iz}^{N|F}$ is defined as the spin current at the NM|FM interface. As in the main text, $d$ is the thickness of the NM layer. An empirical solution of $\delta\vec{\mu}$ is assumed, i.e.

$$\delta\vec{\mu}(z) = \vec{A}\exp[-z/\lambda_N] + \vec{B}\exp[z/\lambda_N] \tag{S3.3}$$

where $\lambda_N$ is the spin diffusion length of the NM layer. The coefficients $\vec{A}$ and $\vec{B}$ are determined by substituting Eq. (S3.3) into Eq. (S3.1) and applying the boundary conditions (Eq. (S3.2)). By doing so, one obtains the spin accumulation at the NM|FM interface

($\delta\bar{\mu}(z=d)$). The current induced effective field scales with the interface spin current $Q_{iz}^{N|F}$, i.e.

$$\Delta H_i = -\frac{\hbar}{2|e|t_F M_S}(\hat{m} \times Q_{iz}^{N|F})_i \qquad (S3.4)$$

$\hbar$ is the Planck constant, $t_F$ and $M_S$ are the thickness and saturation magnetization of the FM layer. The following expression is assumed[6,7] for $Q_{iz}^{N|F}$:

$$Q_{iz}^{N|F} = -\frac{1}{2|e|}\text{Re}[G_{MIX}](\hat{m}\times\hat{m}\times\delta\bar{\mu}(z=d))_i - \frac{1}{2|e|}\text{Im}[G_{MIX}](\hat{m}\times\delta\bar{\mu}(z=d))_i \qquad (S3.5)$$

where $\hat{m}$ is an unit vector representing the magnetization direction and $G_{MIX}$ is the "spin mixing conductance[6-8]" which is a complex number. Substituting $\delta\bar{\mu}(z=d)$ into Eq. (S3.5) gives $Q_{iz}^{N|F}$, which can be plugged into Eq. (S3.4) to obtain the following expression for $\Delta H_i$:

$$\Delta H_x = -\theta_{SH}\frac{\hbar}{2|e|}\frac{J_N}{M_S t_F}\left(1-\frac{1}{\cosh(d/\lambda_N)}\right)\left[\frac{(1+g_r)g_r + g_i^2}{(1+g_r)^2 + g_i^2}\right]\text{sgn}(\hat{m}\times\hat{e}_y)_x \qquad (S3.6)$$

$$\Delta H_y = -\theta_{SH}\frac{\hbar}{2|e|}\frac{J_N}{M_S t_F}\left(1-\frac{1}{\cosh(d/\lambda_N)}\right)\left[\frac{g_i}{(1+g_r)^2 + g_i^2}\right] \qquad (S3.7)$$

where $g_r = \text{Re}[G_{MIX}]\rho_N\lambda_N\coth\left(\frac{d}{\lambda_N}\right)$ and $g_i = \text{Im}[G_{MIX}]\rho_N\lambda_N\coth\left(\frac{d}{\lambda_N}\right)$. Re[$G_{MIX}$] and Im[$G_{MIX}$] represent the real and imaginary parts of $G_{MIX}$. $\Delta H_x$ and $\Delta H_y$ correspond to the damping-like ($\Delta H_L$) and the field-like ($\Delta H_T$) terms, respectively. In deriving Eqs. (S3.6) and (S3.7), the assumption made is that the spin transport at the NM|FM interface is given by Eq. (S3.5) and that contribution from the spins which enter the FM layer is neglected in obtaining $\Delta H_i$ (the effect of spin transport within the FM layer can be included effectively in $G_{MIX}$).

B. Fitting with the model

The thickness and temperature dependencies of the current induced effective field are fitted with Eqs. (S3.6) and (S3.7). As described in the main text, we fix the following parameters to be temperature independent: the FM saturation magnetization ($M_S$=1200 emu/cm$^3$), the NM resistivity ($\rho_N$=395 μΩ·cm), the NM spin diffusion length ($\lambda_N$) and the NM spin Hall angle ($\theta_{SH}$). $\rho_N$ is determined by the underlayer thickness ($d$) dependence of the longitudinal resistance ($R_{XX}$). The real (Re[$G_{MIX}$]) and imaginary (Im[$G_{MIX}$]) parts of the spin mixing conductance at the NM|FM interface are allowed to change with temperature. For a given $\lambda_N$ and $\theta_{SH}$, we fit the underlayer thickness ($d$) dependence of the effective field with Re[$G_{MIX}$] and Im[$G_{MIX}$] as the fitting parameters. The fitting is carried out independently for the effective field measured at 100 K and at room temperature. In order to fit the size and $d$ dependence of the effective field, the allowed values of Re[$G_{MIX}$] and Im[$G_{MIX}$] become quite limited.

Exemplary results of the fitting with different values of $\theta_{SH}$ (and the associated Re[$G_{MIX}$] and Im[$G_{MIX}$] that give the best fit) with a fixed $\lambda_N$ (=2.5 nm) are shown in Fig. S6 for measurement temperatures of 100 K and room temperature (~295 K). The upper two rows of Fig. S6 show the underlayer thickness dependence of the effective field. Values of Re[$G_{MIX}$] and Im[$G_{MIX}$] used at both temperatures are shown in the bottom row with solid symbols. The $d$ dependence of the effective field can be fitted well with different values of $\theta_{SH}$ and $\lambda_N$. $\lambda_N$ is determined by the underlayer thickness at which the effective field saturates, and we find $\lambda_N$ of ~2 to ~3 nm gives the best fit. $\theta_{SH}$ is correlated with the values of Re[$G_{MIX}$] and Im[$G_{MIX}$] and we determine its value by calculating the temperature dependence of the effective field, as described below, and look for the best fit.

The temperature dependence of the effective field (3rd and 4th row of Fig. S6) is calculated using the parameters found above. At each temperature, we substitute the temperature independent parameters ($M_S$, $\rho_N$, $\lambda_N$ and $\theta_{SH}$) and the interpolated values of Re[$G_{MIX}$] and Im[$G_{MIX}$], as shown by the solid lines in the bottom row of Fig. S6, in Eq. (S3.6) and (S3.7). We find that $\lambda_N \sim 2.5$ nm and $\theta_{SH} \sim -0.08$ give the best results. The trend shown in Fig. S6 is similar when $\lambda_N$ varied within the range (~2-3 nm) where the *d* dependence of the effective field can be well reproduced.

**Figure captions**

**Figure S1.** (a-d) Temperature dependence of the normalized resistance $R \cdot w/L$ (a), the anomalous Hall effect resistance $\Delta R_{XY}$ (b), the saturation magnetization $M_S$ (c) and the effective magnetic anisotropy energy $K_{EFF}$ (d) for film structures A: Sub.|$d$ Ta|1 CoFeB|2 MgO|1 Ta and B: Sub.|$d$ TaN|1 CoFeB|2 MgO|1 Ta.

**Figure S2.** Exemplary results of the adiabatic harmonic Hall measurements for film structure A: Sub.|$d$ Ta|1 CoFeB|2 MgO|1 Ta with $d$~0.7 nm. (a,b) In-phase first harmonic ($V_\omega$) voltage (a) and 90° out of phase second harmonic ($V_{2\omega}$) voltage (b) as a function of in-plane external magnetic field for various temperatures. The field direction within the film plane is transverse ($H_T$) to the current flow for (a) and (b) top panel, and along ($H_L$) the current flow for (b) bottom panel. The amplitude of the sinusoidal excitation voltage ($V_{IN}$) is 20 V. Solid and opened symbols correspond to magnetization directed along +z and -z. Parabolic and linear fits to the data are shown by the solid lines. (c-d) Excitation voltage amplitude ($V_{IN}$) dependence of the transverse (c) and longitudinal (d) components of the current induced effective field, obtained using Eq. (S2.1) for various temperatures. Solid and opened symbols correspond to magnetization directed along +z and -z. Linear fit to the data are shown by the solid lines.

**Figure S3.** (a,b) Schematic illustration of the Hall resistance measurements. External field ($H$) of 20-30 kOe is applied to align the magnetization along the field direction. The sample is rotated along the plane shown by the transparent light blue circle. The angle between the film normal and the external field is defined as $\varphi_{OOP}$, whereas the angle between the external field and the current flow is set to be $\varphi_{INP}$. (c,d) Hall resistance $R_{XY}$ as a function of angles $\varphi_{OOP}$ (c) and $\varphi_{INP}$ (d) measured at 100 K. Inset in (d) is a magnified view of the main plot.

Solid line in (c) and (d) shows fit to the data to obtain the anomalous Hall resistance ($\Delta R_A$) and the planar Hall resistances ($\Delta R_P$), respectively. (e) $\Delta R_A$ and $\Delta R_P$ plotted as a function of temperature. (f) Temperature dependence of the ratio $\xi \equiv \frac{\Delta R_P}{\Delta R_A}$. All data presented are for film structure A: Sub.|d Ta|1 CoFeB|2 MgO|1 Ta with $d \sim 1.0$ nm.

**Figure S4.** Ratio of the planar Hall resistance to the anomalous Hall resistance ($\xi \equiv \frac{\Delta R_P}{\Delta R_A}$) plotted as a function of temperature for several devices with different Ta underlayer thicknesses for film structure A: Sub.|d Ta|1 CoFeB|2 MgO|1 Ta.

**Figure S5.** (a,b) Magnitude $|\Delta H / J_N| \equiv \sqrt{(\Delta H_T / J_N)^2 + (\Delta H_L / J_N)^2}$ (a) and the angle with respect to the current flow $\varphi = \arctan\left[\frac{\Delta H_T / J_N}{\Delta H_L / J_N}\right]$ (b) are plotted against the temperature for film structures A: Sub.|d Ta|1 CoFeB|2 MgO|1 Ta and B: Sub.|d TaN|1 CoFeB|2 MgO|1 Ta. The bottom panel of (a) corresponds to a magnified view of the top panel around $|\Delta H|=0$. Solid and open symbols correspond to magnetization pointing along +z and -z, respectively. Oersted field from the current flowing through the underlayer is calculated and its contribution to $\Delta H_T$ is subtracted.

**Figure S6.** (a-j) Underlayer thickness (d) dependence of the transverse $\Delta H_T$ (a-e) and the longitudinal $\Delta H_L$ (f-j) components of the current induced effective field per unit current density ($10^8$ A/cm$^2$) flowing through the TaN underlayer for film structure B: Sub.|d TaN|1 CoFeB|2 MgO|1 Ta. The measurement temperature is 100 K (room temperature, i.e. ~295 K)

for blue (red) symbols. Solid and open symbols correspond to magnetization pointing along +z and -z, respectively. Oersted field from the current flowing through the underlayer is calculated and its contribution to $\Delta H_T$ is subtracted.

(k-t) Temperature dependence of the transverse $\Delta H_T$ (k-o) and longitudinal $\Delta H_L$ (p-t) components of the current induced effective field per unit current density ($10^8$ A/cm$^2$) flowing through the TaN underlayer for film structure B: Sub.|$d$ TaN|1 CoFeB|2 MgO|1 Ta with $d$~8.9 nm. Solid and open symbols correspond to magnetization pointing along +z and -z, respectively. Oersted field from the current flowing through the underlayer is calculated and its contribution to $\Delta H_T$ is subtracted.

Solid lines in (a-t) are calculated using Eqs. (S3.6) and (S3.7) with the following parameters: $M_S$=1200 emu/cm$^3$, $\rho_N$=395 μΩ·cm, $\lambda_N$=2.5 nm and the real and imaginary parts of the spin mixing conductance ($G_{MIX}$), which are shown in (u-y) in the corresponding column. Each column show calculation results with different $\theta_{SH}$: the value used is noted at the top.

(u-y) Temperature dependence of the real and imaginary parts of the spin mixing conductance ($G_{MIX}$) obtained by the fitting of the underlayer thickness dependence shown in (a-j). Linear fit to the data (symbols) is shown by the solid line.

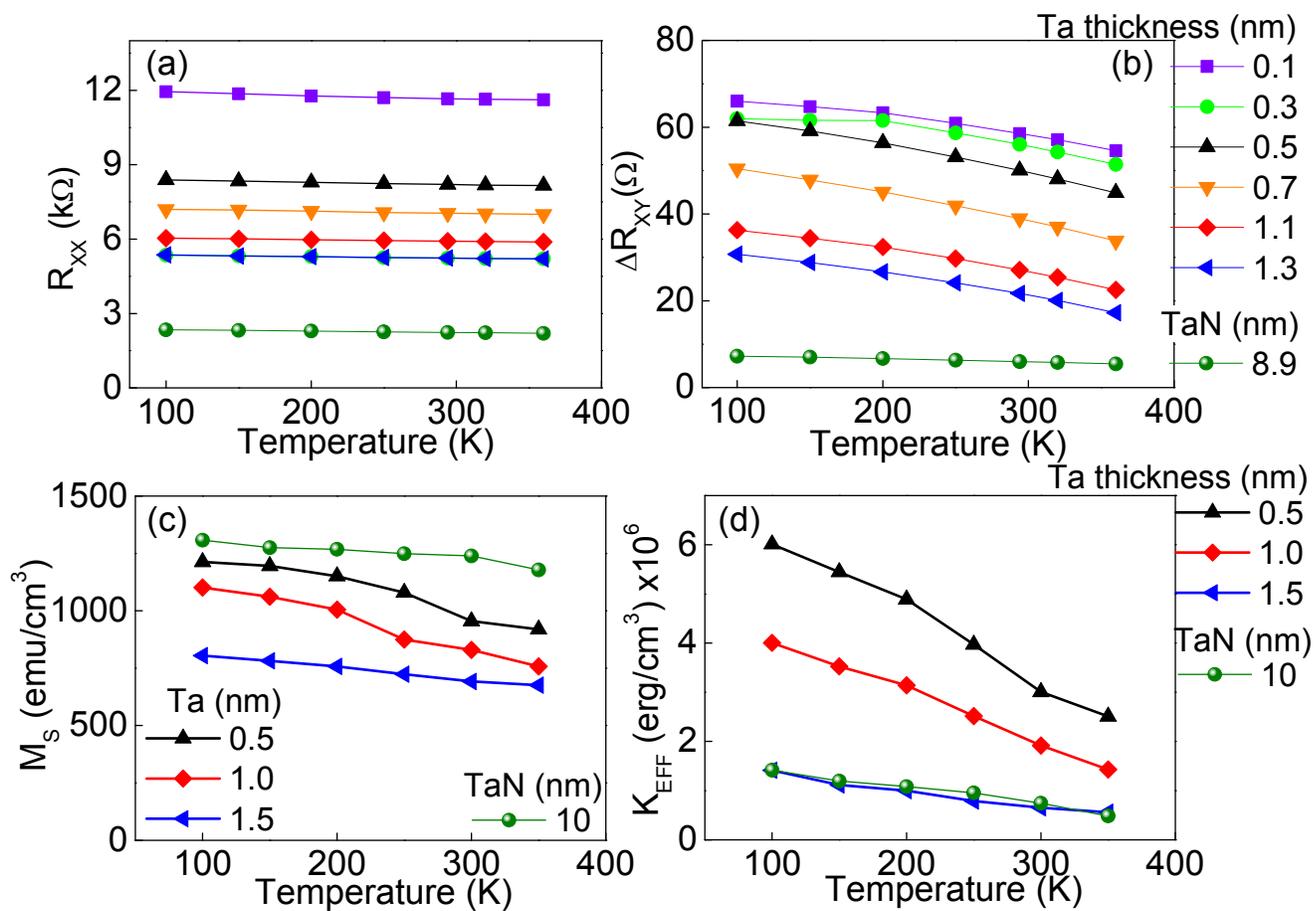

Fig. S1

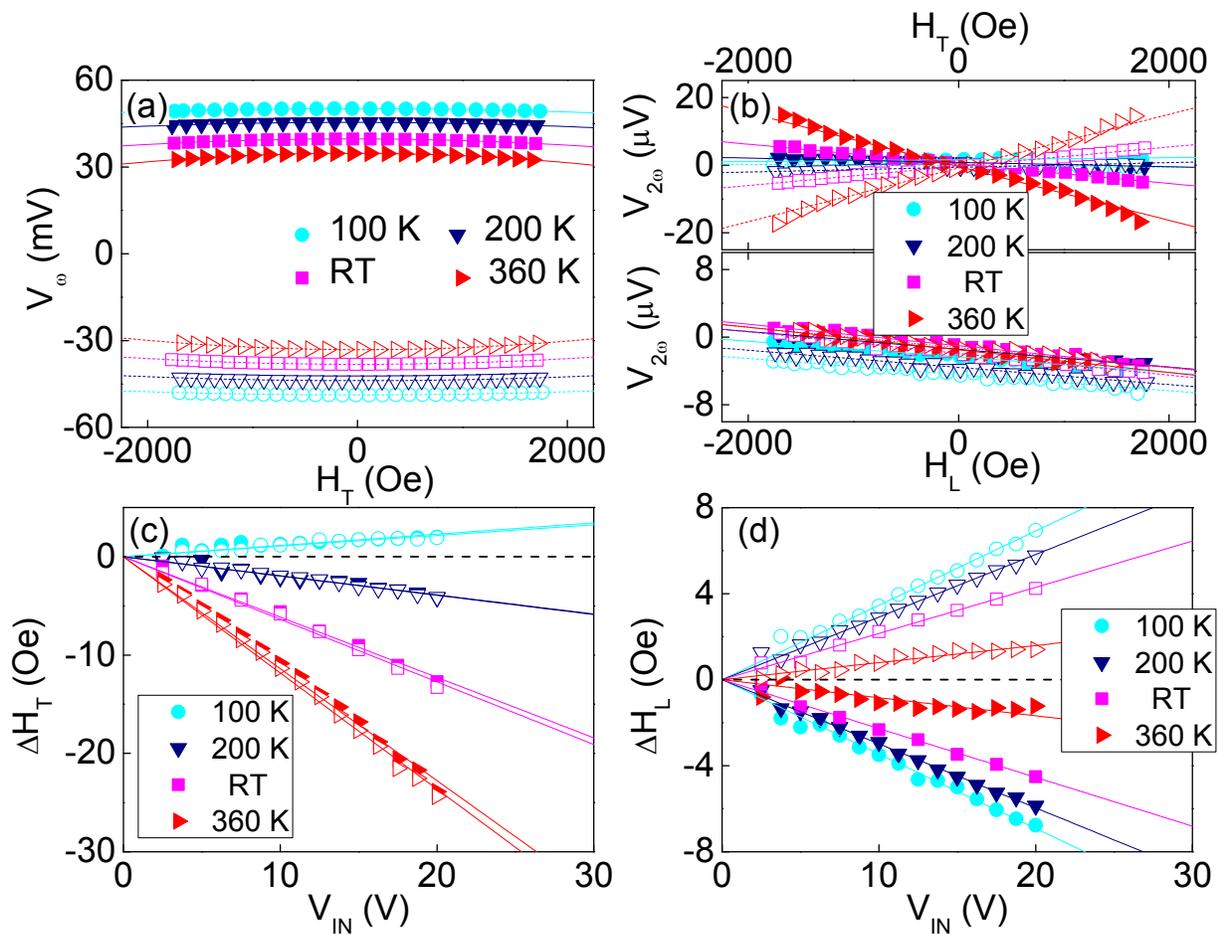

Fig. S2

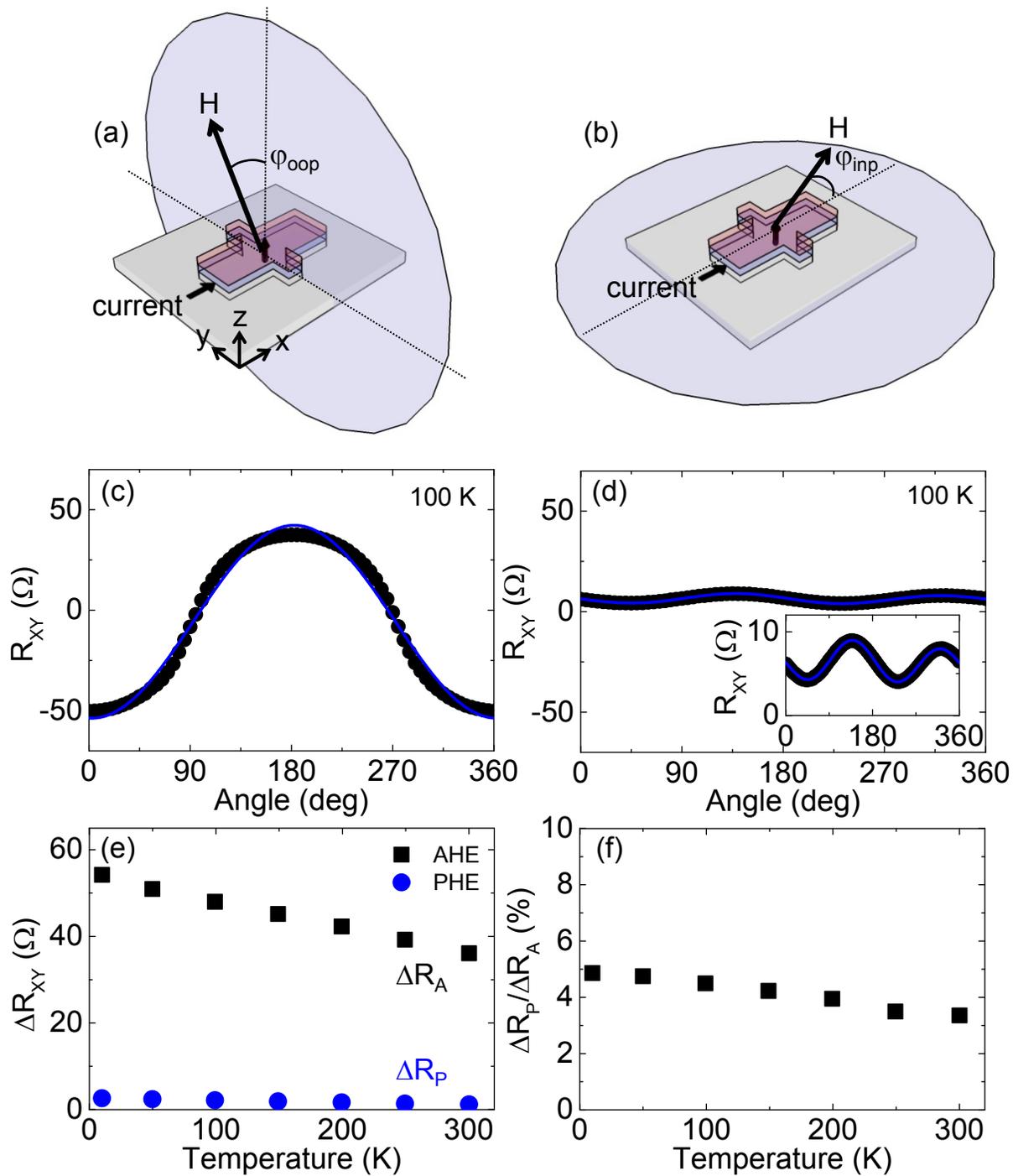

Fig. S3

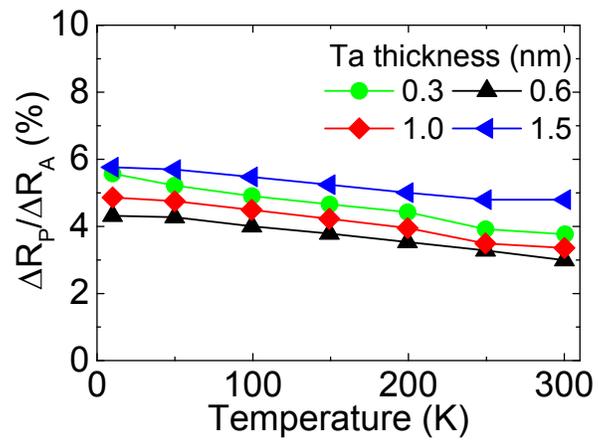

Fig. S4

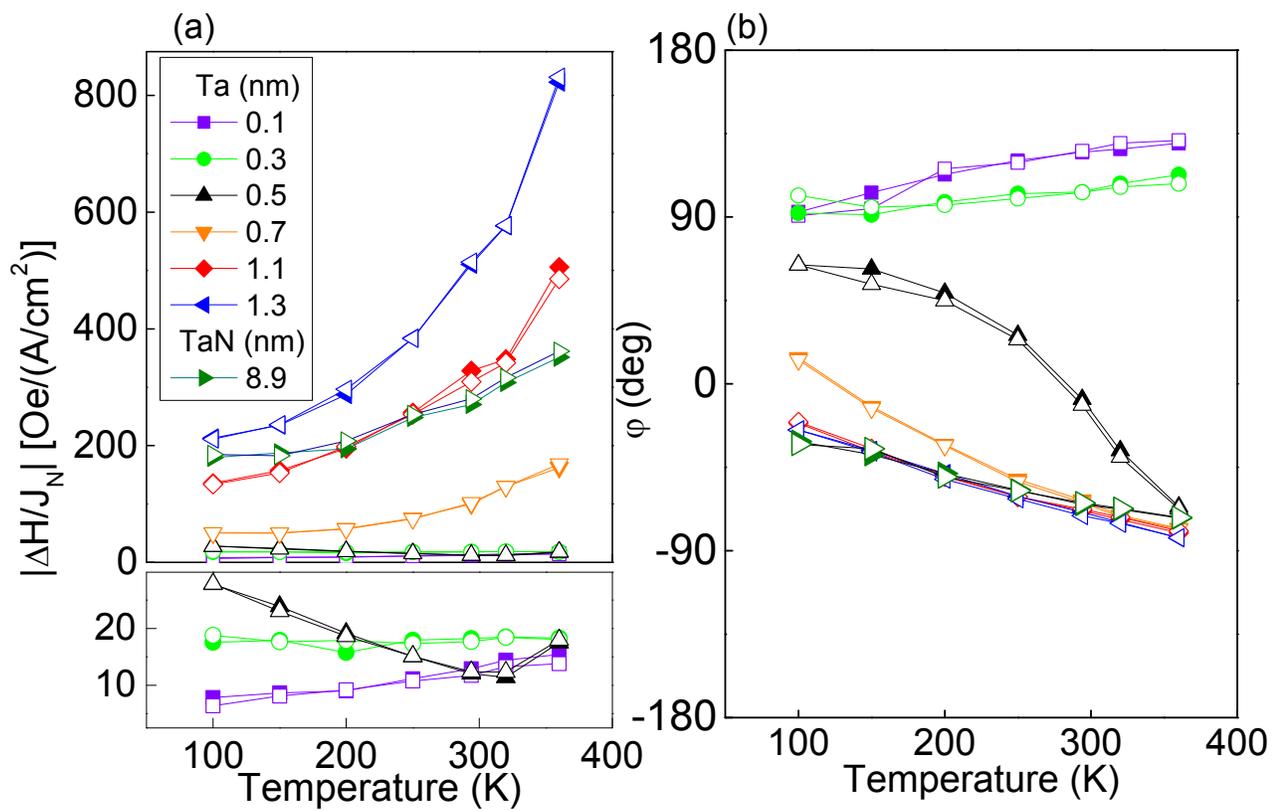

Fig. S5

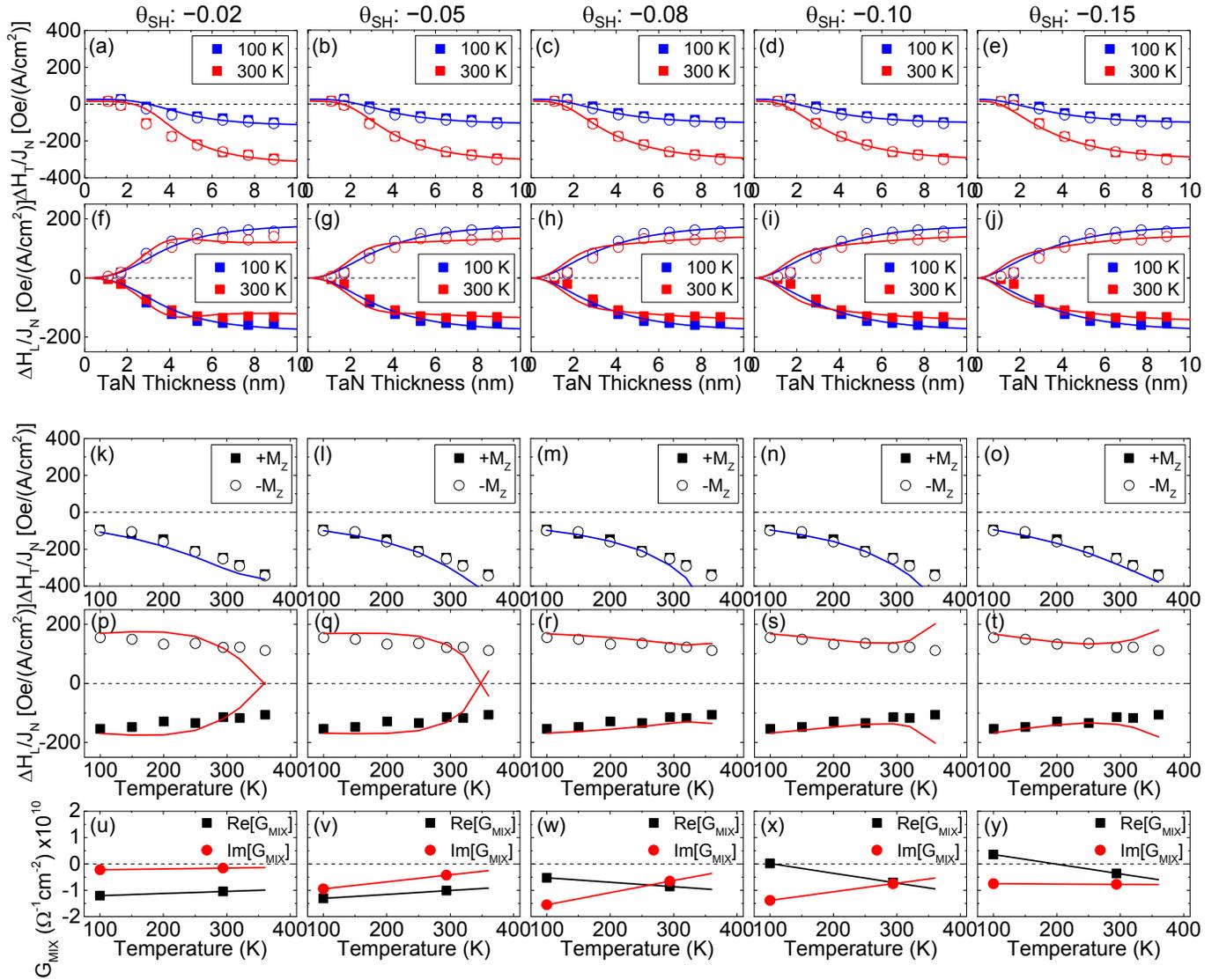

Fig. S6